\documentclass[twocolumn,printnumbers,amsmath,amssymb,showpacs,prl]{revtex4}

\usepackage{graphicx}
\usepackage{color}
\usepackage{ulem}
\usepackage[british]{babel}

\begin{document}
\title{Jamming of packings of frictionless particles with and without shear}

\date{\today}

\author{Wen Zheng}
\email[Email: ]{wenzheng@ustc.edu.cn}
\author{Shiyun Zhang}
\author{Ning Xu}
\email[Email: ]{ningxu@ustc.edu.cn}
\affiliation{Hefei National Laboratory for Physical Sciences at the Microscale, CAS Key Laboratory of Soft Matter Chemistry, and Department of Physics,University of Science and Technology of China, Hefei 230026, People's Republic of China.}

\begin{abstract}
By minimizing the enthalpy of packings of frictionless particles, we obtain jammed solids at desired pressures and hence investigate the jamming transition with and without shear. Typical scaling relations of the jamming transition are recovered in both cases. In contrast to systems without shear, shear-driven jamming transition occurs at a higher packing fraction and the jammed solids are more rigid with an anisotropic force network. Furthermore, by introducing the macro-friction coefficient, we propose an explanation of the packing fraction gap between sheared and non-sheared systems at fixed pressure.

\end{abstract}


\maketitle

\section{I. Introduction}
Jamming transition occurs when a system undergoes the transition from a liquid-like state to a rigid but disordered solid state, which is widely studied in dense disordered systems of colloids, emulsions, foams, and granular materials \cite{liu1998,ohern2003,zheng2015,zheng2017,lin2017}. At zero temperature ($T=0$) and driven by the increase of packing fraction, packings of frictionless particles obtained by fast quench of random configurations jam at a critical-like point called point $J$ associated with a critical packing fraction $\phi_J$ \cite{ohern2003,goodrich2014,heussinger2009,olsson2007,vagberg2011a,wyart2005a,wyart2005b}. Marginally jammed solids right above $\phi_J$ exhibit specific scaling behaviors \cite{bouzid2013,cruz2005,fan2017,goodrich2012,kamien2007,kawasaki2015,wang2015a,liu2014}. Despite recent efforts, a full understanding of the jamming transition is still lacking.

It has been shown that the jamming transition threshold $\phi_J$ is well-defined and corresponds to the random close packing of hard spheres for fast quenched systems \cite{ohern2003,xu2005,zheng2015,liu2014,olsson2007,goodrich2014,liao2018,tong2015}. However, the uniqueness of $\phi_J$ has been questioned by recent studies of the protocol dependence of the jamming transition. It has been illustrated that amorphous, isostatic packings of frictionless spheres can exist over a finite range of packing fractions, with rather weak correlations between structural order and packing fraction, which extend the jamming transition from a jamming point to a jamming line \cite{inagaki2011,ozawa2012,schrech2011,vagberg2011b,vagberg2016,wang2013,berthier2016,charbonneau2017,chaudhuri2010,jin2017,ozawa2017,urbani2017}.

Among the studies of protocol dependence of $\phi_J$, what interests us here is the observations of the $\phi_J$ increase for packings of frictionless particles under load such as shear \cite{liu2014,vagberg2011a,vagberg2016} or self propulsion \cite{liao2018}. A superficial guess of such $\phi_J$ increase may be the presence of a non-zero shear stress or particle activity. It is however interesting to know whether the increase reflects some inherent properties of jammed solids.  A jammed solid has a non-zero pressure and yield stress \cite{liu2014,ciamarra2009,heussinger2009}. Under quasistatic planar shear, it deforms elastically under small shear strains and flows when the shear stress exceeds the yield stress. This is analogical to the stick-slip motion of a pushed object on a rough surface, which implies the existence of an effective macro-friction \cite{boyer2011,bruyn2011,peyneau2008}. By definition, the macro-friction coefficient is equal to the ratio of the yield shear stress to the pressure. We are interested to see whether the $\phi_J$ increase of sheared systems is related to the macro-friction.

Different from previous approaches to minimize the potential energy at fixed packing fraction, here we simply minimize the enthalpy to quickly obtain jammed solids at desired pressures. Because the packing fraction is allowed to vary during minimization, our jammed solids are stable subject to the change of packing fraction \cite{dagois2012}. We explore marginally jammed solids under fixed pressure with and without shear and reproduce well-known critical scalings of the jamming transition. Possibly due to the new minimization method, we obtain a slightly different finite size scaling exponent of $\phi_J$ from previous approaches. With shear, $\phi_J$ is higher than that without shear, as expected, and the jammed solids are more rigid and anisotropic in force network. Furthermore, we find a roughly linear correlation between the macro-friction and the packing fraction gap between sheared and non-sheared solids under the same pressure. We thus propose that the increase of $\phi_J$ under shear results from the macro-friction.

\section{II. Models and Methods}
To avoid crystallization, we put $N/2$ large and $N/2$ small disks with equal mass $m$ into a box with side length $L$.  The diameter ratio of the large to small particles is $1.4$. The interparticle potential is \cite{ohern2003,liu2014,wang2015b,zheng2016}
\begin{equation}
U(r_{ij})=\frac{\epsilon}{\alpha}(1-\frac{r_{ij}}{\sigma_{ij}})^{\alpha}, \label{potential}
\end{equation}
where $r_{ij}$ and $\sigma_{ij}$ are the separation between particles $i$ and $j$ and sum of their radii. We set the units of mass, energy, and length to be particle mass $m$, characteristic energy scale of the potential $\epsilon$, and small particle diameter $\sigma_s$.

To reduce the fuzziness of the packing fraction in quantifying the distance to the jamming transition due to finite size effect, we choose here the pressure as the control parameter. Under constant pressure condition, enthalpy is the right characteristic energy insted of the potential energy. We thus generate static packings at fixed pressure $p$ by applying the fast inertial relaxation engine method \cite{bitzek2006} to minimize the enthalpy $H=U+pV$ of random configurations, where $U=\sum_{ij}U(r_{ij})$ is the sum of potential energy over all pairs of particles and $V=L^d$ is the volume with $d$ being the dimension of space. Without shear, we apply periodic boundary conditions in all direction. To realize quasistatic shear, we successively increase the shear strain $\gamma$ by a small increment $\delta\gamma$, followed by the minimization of the enthalpy under the Lees-Edwards boundary conditions \cite{allen1987}. In this work, we show results for two-dimensional ($d=2$) systems with harmonic repulsion ($\alpha=2$). We vary the system size $N$ from $64$ to $8192$.

\begin{figure}[t]
\vspace{-0 in}
\includegraphics[width=0.48\textwidth]{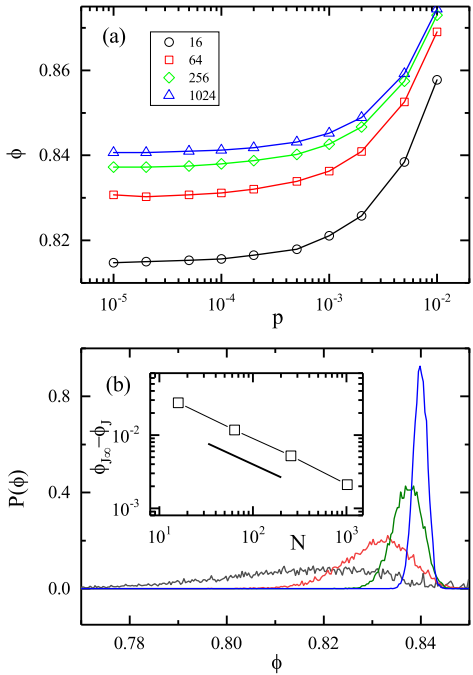}
\vspace{-0.4 in}
\caption{\label{fig:fig1} (color online).
(a) Average packing fraction $\phi$ as a function of pressure $p$ for different system sizes (specified in the legend) in two-dimensions. The solid lines are guides for the eye.
(b) Distributions of $\phi$, $P(\phi)$, at $p=10^{-5}$ for different system sizes. From the left to the right, $N=16, 64, 256$, and $1024$, respectively. Inset: system size $N$ evolution of the gap of the jamming transition thresholds between infinitely large and finite size systems, $\phi_{J\infty}-\phi_J$. The thin line is a guide for the eye. The short thick line has a slope of $-0.616$.
}
\end{figure}

\section{III. Results}
\subsection{A. Finite size scaling of $\phi_J$}
Figure 1(a) shows the pressure dependence of the average packing fraction for different system sizes. Each data point is averaged over thousands of distinct states. When pressure decreases to $0$ (approaching the unjamming transition), the average packing fraction approaches a system size dependent value $\phi_J(N)$. It is well-known that for harmonic repulsion
\begin{equation}
\phi-\phi_J\sim p. \label{pressure}
\end{equation}
We thus obtain $\phi_J(N)$ by fitting curves in Fig.~\ref{fig:fig1} with Eq.~(\ref{pressure}).

In Fig.~1(b), we show the distribution of packing fraction $P(\phi)$ at a small pressure $p=10^{-5}$ for different system sizes. When system size increases, the distribution becomes narrower. The half-height width of the distribution tends to vanish when $N$ approaches $\infty$, implying a well-defined $\phi_J$ in the thermodynamic limit, consistent with previous approaches \cite{ohern2003,liu2014,xu2005}.

Inset of Fig.~\ref{fig:fig1}(b) shows the plot of $\phi_J(N)$, which can be fitted well with the scaling relation
\begin{equation}
\phi_{J\infty}-\phi_J(N)\sim N^{-w}\sim L^{-\nu},
\end{equation}
implying a diverging length scale at $\phi_{J\infty}$ and criticality of the jamming transition at $\phi_{J\infty}$, where $\nu=dw$. Our fitting results in $\phi_{J\infty}=0.8418\pm0.0006$ and $\nu=1.232\pm 0.016$ ($w=0.616\pm0.008$). The scaling exponent $\nu$ is a little smaller than $1.4\pm 0.1$ estimated in previous studies \cite{ohern2003,liu2014,liao2018}. This small difference may come from the difference in minimization. Previous approaches employ minimization at fixed packing fraction, while here we constrain the pressure. As mentioned earlier, our approach here explores stable jammed states subject to the change of packing fraction, which excludes a fraction of states unstable to the packing fraction change but probably counted by constant packing fraction approaches \cite{dagois2012}.

\begin{figure}[b]
\vspace{-0 in}
\includegraphics[width=0.48\textwidth]{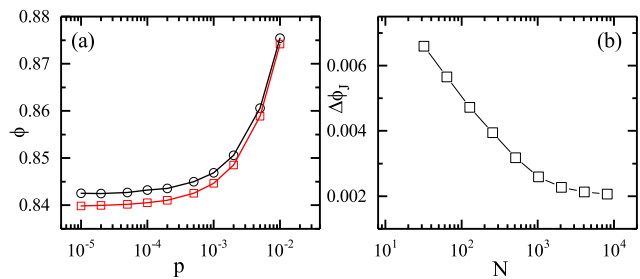}
\vspace{-0.3 in}
\caption{\label{fig:fig2} (color online).
(a) Comparison of the pressure $p$ dependence of the average packing fraction $\phi$ between sheared (circles) and non-sheared(squares) systems. The lines are guides for the eye. (b) System size evolution of the gap of jamming transition thresholds between sheared and non-sheared systems, $\Delta\phi_J$. The line is a guide for the eye.
}
\end{figure}

\subsection{B. Shear induced increase of $\phi_J$}
In Fig.~\ref{fig:fig2}(a), we compare the average packing fraction with and without shear for $N=1024$ systems. At each pressure, we randomly choose tens of initial states and perform quasistatic shear up to $\gamma=10$ with a strain increment $\delta\gamma=10^{-4}$. We then take the average over thousands of sheared states. Figure~\ref{fig:fig2}(a) shows that the average packing fraction for sheared systems is larger than systems without shear and the gap tends to increase approaching $p=0$. By fitting $\phi(p)$ with Eq.~(\ref{pressure}), the jamming transition threshold for sheared systems $\phi_J^s$ is approximately $0.8436$, which is a little bit larger than $0.8403$ for systems without shear. This is consistent with previous observations from simulations under constant packing fraction.

To illustrate that the gap $\Delta\phi_J=\phi_J^s-\phi_J$ is not a finite size effect, in Fig.~\ref{fig:fig2}(b), we show the system size dependence of $\Delta\phi_J$. Although decreasing with increasing $N$, $\Delta\phi_J$ tends to approach a nonzero value in the large system size limit. Note that sheared solids are mostly collected in the ``flowing" regime of the quasistatic shear, where shear stresses fluctuate around the yield stress. The sheared solids thus have nonzero shear stress. One may then question whether the existence of $\Delta\phi_J$ is due to the nonzero shear stress of sheared solids. However, approaching $p=0$, the yield stress approaches zero \cite{liu2014,ciamarra2009,heussinger2009}, so that the shear stress of sheared solids is negligible. Therefore, the presence of shear stress should not be the intrinsic cause of $\Delta\phi_J$. We need to search for other inherent properties of marginally jammed solids responsible to and thus validate the existence of $\Delta\phi_J$, which will be mainly discussed in Sec. III(D).

\begin{figure}[t]
\vspace{-0 in}
\includegraphics[width=0.48\textwidth]{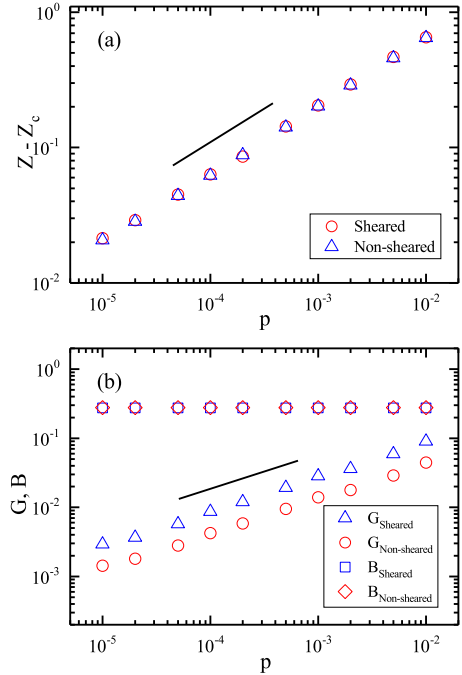}
\vspace{-0.4 in}
\caption{\label{fig:fig3} (color online). Comparison of the pressure $p$ dependence of (a) excess average coordination number per particle $Z-Z_c$ and (b) shear and bulk moduli $G$ and $B$ for systems with and without shear. The short lines have a slope of $1/2$.
}
\end{figure}

\begin{figure*}[t]
\vspace{-0 in}
\includegraphics[width=0.8\textwidth]{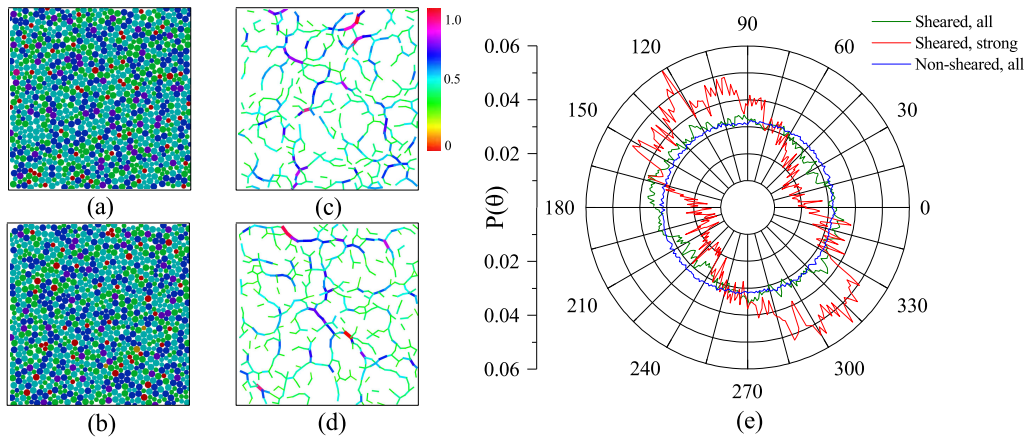}
\vspace{-0.10 in}
\caption{\label{fig:fig4} (color online).Examples of (a, b) configurations and (c, d) structures of strong bonds with and without shear. (a) and (c) are for non-sheared systems, while (b) and (d) are for sheared ones. Different colors in (a) and (b) represent different coordination numbers (purple = 6, blue = 5, cyan = 4, green = 3, and red = 0). The color scale bar calibrates the magnitude of the bond strength. (e) Probability distribution of the angle of the bonds, $P(\theta)$. For sheared systems, distributions for all bonds and strong bonds are both shown. For non-sheared systems, the two distributions are roughly the same, so only the distribution of all bonds is shown here.
}
\end{figure*}

\subsection{C. Properties of sheared marginally jammed solids}
We have shown that $\phi(p)$ for both sheared and non-sheared systems exhibit the same type of scaling. In Fig.~\ref{fig:fig3}, we further compare scalings of the average coordination number per particle $Z$ and elastic moduli with and without shear. The sheared systems still exhibit the well-known jamming scalings: $Z-Z_c\sim p^{1/2}$, $G\sim p^{1/2}$, and $B\sim p^0$, where $Z_c=2d$ is the isostatic value, and $G$ and $B$ are the shear and bulk modulus. Interestingly, the presence of shear does not change the coordination number and bulk modulus even quantitatively, as shown by the nice data collapse in Figs.~\ref{fig:fig3}(a) and (b). However, sheared solids are stiffer subject to shear with a larger shear modulus, as shown in Fig.~\ref{fig:fig3}(b). Then it comes a question what cause the quantitative difference in the shear modulus.

As compared by Figs.~\ref{fig:fig4}(a) and (b), we can hardly tell difference between sheared and non-sheared systems from the configuration and the coordination number distributions. Figures~\ref{fig:fig4}(c) and (d) show spatial distributions of strong bonds with particle interaction larger than the average, from which we can roughly tell that the strong bonds tend to be more anisotropic for sheared systems.

In order to quantify the bond anisotropy, we plot in Fig.~\ref{fig:fig4}(e) the distribution of the angle or direction of bonds. For systems without shear, the angles of all bonds uniformly distribute in $[0^{\circ},360^{\circ}]$, while for sheared systems the distribution seems larger in directions of $135^{\circ}$ and $315^{\circ}$. The anisotropy is pronounced when we plot the angle distribution of strong bonds, as shown in Fig.~\ref{fig:fig4}(e). In contrast, no significant anisotropy emerges in the angle distribution of strong bonds for non-sheared systems (angle distribution of strong bonds is almost identical to that of all bonds for non-sheared systems, which is thus not shown in Fig.~\ref{fig:fig4}(e)). Strong bonds should play more important roles in supporting external loads. The anisotropy exhibited in sheared systems may be the origin of the increase of shear modulus. 

\begin{figure}[b]
\vspace{-0 in}
\includegraphics[width=0.48\textwidth]{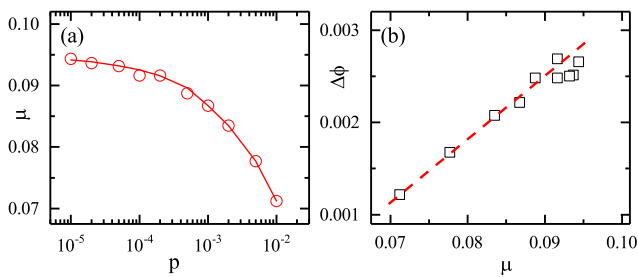}
\vspace{-0.3 in}
\caption{\label{fig:fig5} (color online). (a) Pressure $p$ dependence of the macro-friction coefficient $\mu$ for systems with $N=1024$ particles. The solid line is a power-law fit to the data: $\mu_J-\mu\sim p^{0.5}$ with $\mu_J=0.0952\pm 0.0003$ being the value at the unjamming transition. (b) Correlation between the macro-friction coefficient $\mu$ and the packing fraction gap $\Delta\phi$ between sheared and non-sheared systems. The dashed line is a linear fit.
}
\end{figure}

\subsection{D. Role of macro-friction}
As mentioned earlier, the response of a jammed solid to the quasistatic planar shear is analogical to that of an object on a frictional surface to a push or drag. It is thus natural to expect that the macro-friction is essential to some behaviors of jammed solids under shear, including the increase of the jamming transition threshold discussed in Sec. III(B). Friction originates from atomic interactions between contacting rough surfaces. Although jammed solids are formed by frictionless particles, their disordered structures lead to rough cross sections parallel to the shear direction, which should be the origin of the macro-friction. In Fig.~\ref{fig:fig5}(a), we show the pressure dependence of the macro-friction coefficient $\mu$ of marginally jammed solids. By definition, $\mu=\Sigma_y/p$ with $\Sigma_y$ being the yield shear stress. $\Sigma_y$ is obtained by quasistatically shearing jammed states at pressure $p$ and averaging the shear stress in steady state. With decreasing pressure, $\mu$ increases and approaches a plateau in the small $p$ limit. It has been known that the yield stress of marginally jammed solids is linearly scaled with pressure \cite{liu2014,ciamarra2009,heussinger2009}, it is thus expected that the macro-friction coefficient reaches a constant value near $p=0$.

Interestingly, when we plot the packing fraction gap $\Delta\phi$ between sheared and non-sheared systems against the macro-friction coefficient for all pressures studied, as shown in Fig.~\ref{fig:fig5}(b), $\Delta\phi$ is roughly linear with $\mu$. Approaching $p=0$, both $\Delta\phi$ and $\mu$ reach an almost constant value, so that the plot does not go beyond the maximum $\mu$ around $0.0952$.

The sheared solids concerned here have nonzero shear stresses fluctuating around the yield stress. If the macro-friction were zero, the yield stress would vanish and applying shear deformation would not affect jammed solids. When $\mu$ increases, the difference induced by shear would grow. Therefore, it is expected that $\Delta\phi$ increases with increasing $\mu$. Approaching the unjamming transition, both the pressure and yield stress approach zero, but their ratio remains almost constant. It is thus the macro-friction but not the shear stress that plays the key role in determining the response ability of marginally jammed solids to shear, particularly non-vanishing $\Delta\phi_J$ observed here.

\section{IV. Discussion and conclusions}
By performing the minimization of enthalpy, we obtain marginally jammed solids under constant pressure and compare the jamming transition and properties of marginally jammed solids with and without shear. The new minimization approach excludes unstable states subject to the change of packing fraction, but the well-known scaling relations obtained by potential energy minimization under constant packing fraction are still recovered, except that here we achieve a slightly smaller finite size scaling exponent of the jamming transition threshold. When shear is applied, the jamming transition is pushed to a higher packing fraction and the jammed solids are more rigid with a larger shear modulus. Our analysis attributes the increase of the jamming transition threshold to the existence of nonzero macro-friction and the enhancement of the rigidity to the anisotropic angle distribution of strong bonds.

The macro-friction comes from disordered structures of jammed solids. We have proposed to use it to explain the shear induced increase of the jamming transition threshold. Moreover, previous studies have shown that the yield stress of marginally jammed solids decreases with increasing system size \cite{liu2014}. By definition, the macro-friction also decreases when system size increases. From our analysis, the gap of the jamming transition thresholds between sheared and non-sheared systems should decrease as well, which is exactly what we observe in Fig.~\ref{fig:fig2}(b). All these evidences suggest that the macro-friction is an important inherent property to determine system's response to shear, whose role in characterizing jamming transition and even the order-disorder transition \cite{tong2015} needs to be reevaluated. Note that here we consider frictionless particles. If particles themselves are frictional, it is interesting to know in follow-up studies how the microscopic friction affects the macro-friction and behaviors of jammed states under shear.

\section{Acknowledgments}
We are grateful to Atsushi Ikeda and Jie Lin for helpful discussions. This work is supported by National Natural Science Foundation of China Grants No.~11702289, No.~11734014, and No.~11574278, and the Anhui Provincial Natural Science Foundation Grant No.~1708085QA07. We thank the Supercomputing Center of University of Science and Technology of China for the computer time.

\end{document}